\documentclass[12pt,preprint]{aastex}

\shorttitle{Are Young Jupiters Faint?}
\shortauthors{Marley et al.}

\begin{document}


\title{On the Luminosity of Young Jupiters}


\author{Mark S. Marley}
\affil{NASA Ames Research Center; Mail Stop 245-3; Moffett Field CA 94035}
\email{Mark.S.Marley@NASA.gov}

\author{Jonathan J. Fortney}
\affil{Spitzer Fellow, NASA Ames Research Center; Mail Stop 245-3; Moffett Field CA 94035 and SETI Institute, 515 N. Whisman Rd., Mountain View, CA 94043}
\email{jfortney@mail.arc.nasa.gov}

\author{Olenka Hubickyj}
\affil{NASA Ames Research Center; Mail Stop 245-3; Moffett Field CA 94035}

\author{Peter Bodenheimer}
\affil{UCO/Lick Observatory, University of California, Santa Cruz CA 95064}
\and

\author{Jack J. Lissauer}
\affil{NASA Ames Research Center; Mail Stop 245-3; Moffett Field CA 94035}



\begin{abstract}
Traditional thermal evolution models of giant planets employ arbitrary initial conditions selected more for computational expediency than physical accuracy.  Since the initial
conditions are eventually forgotten by the evolving planet, this approach is valid for mature planets, if not young ones.  To explore the evolution at young ages
of jovian mass planets we have employed model planets created by one implementation of the core
accretion mechanism as initial conditions for evolutionary calculations.  The luminosities and early cooling rates of young planets are highly sensitive to their internal entropies, which depend on the formation mechanism and are highly model dependent.  As a result of the accretion shock through which most of the planetary mass is processed, we find lower initial internal entropies than commonly assumed in published evolution tracks. Consequently  young jovian planets are smaller, cooler, and several to 100 times less luminous than predicted by earlier models.  Furthermore the time interval during which the young jupiters are fainter than expected depends on the mass of planet.  Jupiter mass planets ($1\,\rm M_J$) align with the conventional model luminosity in as little at 20 million years, but $10\,\rm M_J$ planets can take up to 1 billion years to match 
commonly cited luminosities, given our implementation of  the core accretion mechanism.  If our assumptions, especially including our treatment of the accretion shock, are correct and if extrasolar jovian planets indeed form with low entropy, then young jovian planets are substantially fainter at young ages than currently believed.  Furthermore early evolution tracks should be regarded as uncertain for much longer than the commonly quoted $10^6$ years.  These results have important consequences both for detection strategies and for assigning masses to young jovian planets based on observed luminosities.

\end{abstract}





\section{Introduction}

In the past decade, a number of nearby star associations have been recognized as being quite young, less than $10\,\rm Myr$ old (e.g. IC 348, TW Hydrae, MBM12, $\eta$ Cha \citep{Lada95, Webb99, Luhman04}).  Such associations are likely well stocked with recently formed, presumably
bright giant planets that should in principle be easy prey for a variety of planet detection technologies.  Planning for the hunt, however, requires
knowledge of the expected luminosity of young giant planets as a function of time since their formation, particularly at young ages when they are presumably easy game.

While models of   the luminosity evolution of giant planets have a long pedigree \citep[e.g.,][]{Grossman72, Graboske75}, the early work focused on the evolution of the solar system giants, attempting to explain their current luminosity at an age of 4.5 Gyr. Since planets lose memory of their initial conditions over time, initial conditions were 
selected more for computational convenience than for accuracy.  Many improvements have subsequently been made to the models, particularly in the characterization of jovian atmospheres at various effective temperatures, although essentially the same initial conditions are still employed \citep{Hubbard80a, Burrows97, Chabrier00b}.  

The standard evolution model begins with a hydrogen-helium sphere having a large radius, high internal
entropy, and large effective temperature.  Such an object is not necessarily one that would be the result of any particular planet formation model.  This model planet is allowed to radiate and cool over time.  Since 
there have been no detections yet of young planets with measured masses, the applicability of this initial
condition is untested, although there are data from more massive objects.  A pair of eclipsing, young ($\sim 1\,\rm Myr$) brown dwarfs with known dynamical masses (57 and 36 Jupiter masses ($\rm M_J$)) indeed  have
radii exceeding five times that of Jupiter \citep{Stassun06}, confirming that young massive brown dwarfs, at least, are in fact large and hot.  But giant planets that formed in a disk around a primary star, or even isolated planet-mass objects, may have experienced very different initial conditions.

Evolution tracks computed in the usual way--even though they do not
necessarily reflect any particular planet formation theory--have been used to evaluate detection strategies for true giant planets orbiting solar type stars \citep[e.g.,][]{Burrows05}, and they have been used to characterize isolated, very low mass brown dwarfs.
\citet{Chauvin04}, for example, reported on the detection of a faint companion to the M8 brown dwarf \object{2MASSWJ1207334-393254} (hereafter 2MASS1207) in TW Hydrae with an estimated age of 8 Myr.  Using its observed luminosity and applying published evolution tracks, they estimated a mass of just $5\,\rm M_J$ for the mass of the companion. 

The early modelers certainly did not foresee that direct detections of putative young planets would be compared against the models at exceptionally young ages, at times when the model planet may not yet have forgotten its hot start. \citet{stevenson82} wrote that evolution calculations ``...cannot be expected to provide accurate information on the first $10^5$--$10^8$ years of evolution because of the artificiality of an initially adiabatic, homologously contracting{\footnote{See \citet{Stahler88} for a discussion of homologous contraction.}} state.''  More recently, \citet{Baraffe02} examined the uncertainties in evolution tracks of brown dwarfs at young ages and cautioned about the applicability of evolution models at ages less than a few million years, on the lower end of Stevenson's uncertainty range. \citet{Wuchterl05} has also  expressed concern that standard evolution models do not capture the early evolution correctly.  Given the clear imperatives to interpret
observations of young, low mass objects and to plan for future direct detections of giant planets formed in orbit about solar type stars, there is a need to connect models of giant planet formation to giant planet evolution.

Our goals here are both to help fill the void in physically plausible models of extrasolar giant planets (EGPs) at young ages and to better
quantify the age beyond which the evolution models are robust and applicable.  We aim to understand whether or not the current generation of evolution
models can reliably predict the luminosity of giant planets at young ages and, if not, then define the age beyond which current models are reliable.
Instead of using an arbitrary starting condition, we employ planets formed by one implementation of the core accretion model.  In this scenario gas giant planets form by rapid accretion of gas onto a solid core that grew by accretion of planetesimals in the nebula.  This mechanism is one of two competing scenarios of gas giant formation, the other being the gas instability model by which giants form from a local disk instability \citep{Boss98} that creates  
a self-gravitating clump of gas.  The planet resulting from such a clump could also be used as the starting point of an evolutionary calculation (see \citet{Bodenheimer74, Bodenheimer76, Bodenheimer80}), but we choose here to focus solely on the core accretion mechanism as it currently seems the more promising mechanism for explaining the formation of the giant planets (see \citet{Lissauer06} for a review).

For specificity, we rely upon the implementation by \citet{Hubickyj05} of the core accretion mechanism.   By necessity, their work  makes a host of assumptions that ultimately affect the properties of newly born giant planets.  As we will demonstrate, following the end of accretion this model predicts that giant planets are substantially
fainter than standard evolution models.  While we believe that this conclusion is secure, we stress that the precise numerical value of the post-accretion luminosity depends
upon the particular assumptions employed by \citet{Hubickyj05}.  Therefore, we first briefly review this model and highlight the 
assumptions upon which the work rests in Section 2.  We describe our method of evolving these model planets over time 
in Section 3 and compare our results with 
standard giant planet evolution models. We find in Section 4 that the initial conditions influence subsequent planetary evolution for longer than generally appreciated and that planets formed by the  \citet{Hubickyj05} recipe for core accretion are substantially fainter than standard models have previously predicted.  We end by cautioning those who wish to rely upon evolution models to characterize detected young giant planets
that may have grown by the core accretion mechanism would be wise to be judicious in their estimation of the model-dependent uncertainties.

\section{Accretion}
The core accretion model describing the formation of giant planets \citep{Mizuno80, Bodenheimer86, Pollack96}   suggests that any planet that becomes more massive than about $10\,\rm M_\earth$ (Earth masses) while residing within a gas-rich protoplanetary nebula should accrete a gaseous envelope.  This leads to the expectation that massive planets acquire a thick envelope of roughly nebular composition surrounding a denser core of rock and ice.  In this section, we briefly review the particular implementation of this model by \citet{Pollack96} and collaborators \citep{Bodenheimer00, Hubickyj05} and inform the reader of important model assumptions.  The newborn planets
delivered by this modeling approach are then used as  initial conditions to our own evolution calculations as described in Section 3.

\subsection{Model Overview}

 \citet{Bodenheimer00} describe the core accretion-gas capture process.    The stages described below are chosen for clarity, and do not match the  accretion phases as defined in  \citet{Bodenheimer00}.  Each stage is keyed to Figure 1, which illustrates the luminosity evolution of an accreting $1\,\rm M_J$ planet.

\noindent{1. Dust particles in the solar nebula form planetesimals that accrete into a solid core surrounded by a very low-mass
gaseous envelope. During runaway solid accretion the gas accretion rate is much lower than that
of solids. As the solid material in the feeding zone is depleted,
the solid mass accretion rate and consequently the luminosity fall.  At the end of this stage, most of the mass of the planet consists of solids.  }

\noindent{2. The protoplanet continues to grow as the gas accretion rate steadily increases, eventually
exceeding the solids accretion rate.  The mass of both components grow until  the core and envelope masses
become equal.}

\noindent{3. Runaway gas accretion occurs and the protoplanet grows
rapidly. The evolution up to this point is referred to as
the {\it nebular stage}, because the outer boundary of the protoplanetary
envelope is in contact with the solar nebula
and the density and temperature at this interface merge with nebular values.  During this stage, the nebula is assumed to provide the planet with enough mass that the planet always fills its effective accretion radius, which is almost as large as the radius of its Hill sphere \citep{Bodenheimer00}\footnote{The core continues to grow during this time as large planetesimals are accreted.  Final core masses range from 17 to $19\,\rm M_\earth$ for our 1 and $10\,\rm M_J$ models, respectively.}. }

\noindent{4. As the planet grows, its hunger for gas increases, however the rate of gas consumption is limited
to the rate at which the nebula can transport gas to the vicinity
of the planet. Subsequently, the region of the protoplanet in hydrostatic equilibrium
contracts inside the effective accretion
radius (which at this time is close to that of the Hill sphere), and gas accretes hydrodynamically onto
the planet.  Note that the $1\,\rm M_J$ accretion model we employ is
that of \citet{Hubickyj05}.  For the more massive planets, we
allow this $1\,\rm M_J$ model to spend more time in this phase, with a gas accretion rate  $\dot M_{\rm gas} \approx 10^{-2}\,\rm M_\earth \, yr^{-1}$, until the planet reaches its final target mass.  Accretion is stopped by either the opening of a gap in the
disk as a consequence of accretion, the tidal effect of
the planet, by dissipation of the nebula, or some combination of all three.
For computational stability, $\dot M_{\rm gas}$ is linearly decreased from the limiting rate to zero over a time period of
$3.5\times10^{4} {M\over{1\,\rm M_J}}\,\rm yr$, where $M$ is the mass of the planet. }

During this time of rapid gas accretion, the accreting gas is assumed
to fall from the Hill sphere radius down to the surface of the planet.  It arrives at a shock interface where almost all of the
initial gravitational potential energy of the gas is radiated away upwards, as occurs for accreting stars \citep{stahler80}.  This produces a rapid increase in luminosity and the planet briefly shines quite brightly.  {\em Crucial to the problem at
hand is that  the gas arrives at the surface of the planet having radiated away most of its gravitational potential energy  and 
initial specific entropy and having equilibrated with the local thermal radiation field.}  

\noindent{5. Once accretion
stops, the planet enters the isolation stage. During this stage the planet contracts and cools to the present state at constant
mass. The details of our calculation of the planet's evolution are presented in the Section 3.}

\subsection{Important Model Assumptions}

      Detailed descriptions of the procedure and assumptions 
       that enter into the implementation 
       of the core accretion model  are reported in \citet{Pollack96}, \citet{Bodenheimer00}, and \citet{Hubickyj05}.  For this study we prepared core accretion models for masses of 2, 4, 6, 8, and  $10\,\rm M_J$ by starting with the $1\,\rm M_J$ baseline case (denoted $10\rm L \infty$) from Hubickyj et al. (2005).
       For simplicity, we assumed Stages 1 through 3 to be identical for all 
       planetary masses  and only the total duration of Stage 4 controls the final mass of the planet.
  The special cases (e.g., the high nebular temperature case)
       are new models, computed in the same way as the others, except for
       the one changed parameter.
       
       The protoplanet grows as a lone embryo in a solar nebula of a 
       temperature $T_{\rm neb} = 150\,\rm K$ and density $\rho_{\rm neb} = 5\times 10^{-11}\,\rm g\,cm^{-3}$ with
       a protosolar hydrogen to helium ratio.  The planetesimal feeding zone, which is assumed to be an annulus
       extending to a radial distance of about 4 Hill-sphere radii on 
       either side of the planet's orbit,
       grows as the planet gains mass.  It is assumed that gas from the 
       surrounding solar nebula flows freely, up to a limiting rate, into the evacuated volume.

       The atmospheric boundary condition for the entire core accretion 
       evolution relies upon a gray atmosphere computed with Rosseland 
       mean opacities.  This mean opacity is controlled by the assumed 
       grain number density and size distribution of particles arriving 
       from the nebula.   We employ the `L' models of Hubickyj et al. 
       (2005) in which the opacity due to grains is 2\% of 
       the interstellar grain opacity.  This  grain opacity is
       in agreement with computations by  Podolak (2003) that 
       indicate that when the grains enter the protoplanetary envelope, 
       they coagulate and settle out quickly into warmer regions where 
       they are destroyed, resulting in actual opacities in this low
       temperature region far smaller than interstellar values. Nonetheless,
       throughout the relevant effective temperature range grain opacity dominates the pure gaseous opacity.

       During stage 3, the gas accretion rate increases very quickly.  We 
       limit this increase to the rate at which the solar nebular can 
       supply gas to the planet.  Typical protoplanetary nebula models
       state that the mass  transfer rate, caused by viscous effects, is 
       about $1\times 10^{-2}\,\rm M_\earth \, yr^{-1}$.  When  this limiting rate is reached, the 
       planet contracts inside its accretion radius, but is still assumed to 
       be in hydrostatic equilibrium. The accreting gas is delivered
       hydrodynamically onto the planet at near free-fall velocities.  This 
       hydrodynamic arrival of nebular gas creates a shock at the upper
       boundary of the  planet's atmosphere.  Regardless of its thermal 
       energy, gas is presumed to be delivered homogeneously over the 
       entire surface of the planet.  In fact, since the gas is accreted 
       from a circumplanetary disk, the morphology of  accretion may be 
       quite different.  We do not consider such issues here, although they could be of some importance.

{\em The treatment of mass and energy delivered through this shock \citep{stahler80} is the single most important influence on the final thermal state of the planet.}  The gas is assumed to fall from the radius of the Hill sphere onto the shock, which lies at the upper boundary of the planetary atmosphere.  To explore sensitivity to the thermal state of the pre-accreted gas, we computed  a $2\,\rm M_J$ model with twice the assumed temperature for the nebular gas (300 instead of 150 K).

The precise luminosities expected from core accretion depend upon the assumed profile of the accretion rate, which 
is highly uncertain and in turn rests on assumptions about the ability of the nebula to supply gas to the planet.  For these reasons,
we do not place high confidence in the quantitative comparison of details of the early luminosity evolution between the various core accretion model masses.  To explore the sensitivity of the evolution to the accretion rate, we computed models for $2\,\rm M_J$ with 1/10 and 10 times the baseline gas accretion rate, $\dot M_{\rm gas}$.   We do, however,
regard with confidence  the very large qualitative difference between planets that begin the isolation stage as relatively cool, low entropy 
objects and those which begin with the high entropy hot start as described in Section 3.3.

\section{Evolution}

We evolved each of  the core accretion models to understand their thermal evolution subsequent to their formation. In principle, the core accretion planet formation code could be used to follow the subsequent cooling of each model planet.  However, the grain-laden atmospheres that are incorporated into the formation calculation are not relevant after accretion ceases, when only relatively condensate-free gasses, mixed upwards from deeper in the atmosphere, are relevant.  Thus, once the planet is fully formed we switch over to our fully non-gray EGP/brown dwarf atmosphere code in order to follow the planet's subsequent evolution. In this section we explain precisely how we compute this evolution as well as our `hot start' evolution to which we compare results.

\subsection{Initial Conditions for Cooling}
Our planetary evolution code has previously been applied to the cooling and contraction of Jupiter and Saturn \citep{FH03}, cool EGPs \citep{FH04}, and hot Jupiters \citep{Fortney06}.  To begin the calculation, we employ the envelope model at the termination of accretion from the core accretion code as the starting model for the subsequent evolution calculation.  In the predominantly H/He envelope, we sample the deep convective interior to determine the specific entropy ($S$) of the planetary adiabat.  We then construct a model planet with this same specific entropy (shown in Figure 2) for the start of the evolution phase.  This ensures that the envelope has the same pressure/temperature/density profile at this boundary. Both the formation and evolution codes use the H/He EOS of \citet{SCVH} with $Y$=0.243.  One structural change that we do make is in the core. While the formation code assumes a uniform core density of 3.2 g cm$^{-3}$, the evolution code uses the ANEOS equation of state for olivine \citep{ANEOS}, which allows for the expected significant compression of the core material.  The core mass remains the same, but the core radius is substantially smaller in the evolution phase of the calculation\footnote{We do not account for the gravitational potential energy that would be released if the core were to actually shrink.}.  However, the exact structure of the core has little effect on the evolution, especially for masses $\ge 2\,\rm M_J$, because the core is but  a small fraction of the planet's mass.

Our transition from formation to subsequent evolution involves a change in the outer boundary condition as well.  During the formation phase, the outer boundary is appropriate for a planet embedded in the nebula, but during the evolution phase, the outer boundary condition is that of an isolated planet.

\subsection{Atmospheric Boundary Condition}

We employ a grid of non-gray radiative-convective atmosphere models to compute the evolution of giant planets and brown dwarfs. This grid relates the specific entropy of the adiabatic planetary interior ($S$) and surface gravity ($g$) to the planetary atmosphere's effective temperature, $T_{\rm eff}$, through a relation $T_{\rm eff} = f(g,S)$. (Here $S$ is parameterized as $T_{10}$, the temperature the adiabat would have at a pressure of 10 bar.) \citet{Saumon06} have computed a cloud-free grid of atmospheres from $T_{\rm eff}=500$--$2400\,\rm K$ and $\log g=3.5$--$5.5\,\rm cm\,sec^{-2}$.  We have computed $\sim$75 additional model atmospheres to extend this grid down to $T_{\rm eff}$=90 K and $\log g=1.0$, to cover the lower effective temperatures and gravities necessary to study the evolution of 1 to $10 \,\rm M_J$ planets.  The assumption of cloud-free atmospheres is valid here, because, as we show below, effective temperatures for these planets cluster around 500--800 K at young ages, while water cloud condensation should not begin until $T_{\rm eff} < 500$ K \citep{Burrows03b}. 

The atmosphere code has previously been implemented for a variety of planetary and substellar objects.  Applications include the generation of pressure-temperature (\emph{P--T}) profiles and spectra for Titan \citep{Mckay89}, brown dwarfs \citep{Marley96, Burrows97, Marley02, Saumon06}, Uranus \citep{MM99}, and hot Jupiters \citep{Fortney05,Fortney06}.  The radiative transfer solving scheme is described in \citet{Toon89}.  We use the elemental abundance data of \citet{Lodders03} and compute chemical equilibrium compositions following \citet{Fegley94}, \citet{Lodders02}, and \citet{Lodders02b}.  The large and constantly updated opacity database is described in R.~S.~Freedman and K.~Lodders (2007, in prep.).

\subsection{Hot Start Models}

To compare the evolution calculations employing the core accretion models as the initial condition to the type of evolution models primarily represented in the literature, we computed a second set of models employing what we term a ``hot start.''  These models assume that the planet, at all ages, has reached its final mass and possesses a fully adiabatic interior.  An initial model is chosen with a high specific entropy adiabat (see Figure 2), corresponding to high internal temperatures.  Our choices for initial entropy are very similar ($<10$\% difference) to
those employed by \citet{Burrows97}.   The heat extracted from the planet's interior per unit mass is given by:
\begin{equation}
\label{S}
\frac{\partial L}{\partial m} = -T\frac{\partial S}{\partial t},
\end{equation}
where $L$ is the planet's intrinsic luminosity, $T$ is the temperature of a mass shell, $S$ is the specific entropy of a mass shell, and $t$ is the time.  At the start of an evolutionary sequence $L$ is large, so the time steps $\partial t$ are consequently small.  We note that cooling curves can be constructed back to an arbitrarily young age with this procedure, although authors such as \citet{Burrows97}, \citet{Chabrier00b}, \citet{Baraffe03}, who utilize this technique because of its convenience, typically only plot evolution for ages $> 10^6$ years, with the notion that the evolution at younger ages with this formalism probably does not correspond to reality.  With our own ``hot start'' models, we reproduce well the results of \citet{Burrows97} and \citet{Baraffe03} for the early evolution of 1 to $10\,\rm M_J$ planets; those authors used cloud-free atmosphere models similar to the ones we use here.  

The thermal time scale, $\tau$, of the evolving planet is roughly given by the Kelvin-Helmholtz time scale, 
\begin{equation}
\label{KH}
\tau \approx \frac{GM^2}{RL},
\end{equation}
where $G$ is the gravitational constant, $M$ is the mass of the planet, and $R$ is the radius of the planet.  Since the hot start planets have much larger initial $L$ and $R$, their initial cooling rate is correspondingly faster than that of the core accretion planets.  
Since both the hot start and core accretion evolution tracks utilize the same atmospheric boundary conditions and model approach, any difference between the two must be attributable to the different initial conditions.

\section{Discussion}
\subsection{Luminosity, Mass, and Radius of Young Planets}

By combining our extension of the protoplanetary accretion calculation described in \citet{Hubickyj05} with our calculation of the
subsequent evolution, we produce models of the evolution of luminosity, radius, effective temperature ($T_{\rm eff}$)
of the planet with time.  Our results are shown in Figures 3 and 4 which compare the core accretion evolutionary tracks to the conventional `hot-start' scenario.  

At very young ages, while a planet is still forming, the luminosity is of course far lower than the hot-start case where the planet is assumed to instantaneously form at time $t=0$.
During runaway gas accretion (around 2.5 Myr), the luminosity, which is almost entirely derived
from the accretion shock, peaks in the range of 0.1 to $0.01\,\rm L_\odot$ (solar luminosity), although the precise value is highly dependent on the assumed limiting gas accretion rate and the shock physics.  A planet caught during this time period would be brighter than at any other time during its evolution.

As gas accretion is turned off in the core accretion case, the luminosity rapidly collapses to between $10^{-5}$ to $10^{-6}\,\rm L_\odot$, depending upon the mass.  At this point, the more massive planets have lower entropies (Figure 2), because a proportionately greater amount of their mass has passed through the shock and arrives with low entropy.
As a result,  post-accretion luminosity decreases with increasing mass (Figure 3), a result that is entirely
a consequence of our treatment of the accretion shock.  With the lowest gravity, the $1\,\rm M_J$ planet has the largest radius (Figure 3) and the highest post-formation luminosity.  

The `hot-start' models begin with arbitrarily large initial luminosities, greater than $10^{-4}\,\rm L_\odot$,
 that expeditiously decay away.  Since these planets start fully formed, the choice of time $t=0$ for comparison
 to the core accretion models is somewhat arbitrary.  In Figure 3, we equate the time of the first hot-start model
 to time $t=0$ for the core accretion model.  This allows the hot-start models a 2 to 3 million year `head start' in their cooling and
 consequently minimizes the difference from the core accretion predicted luminosity.  Nevertheless,
 with the sole exception of the $1\,\rm\, M_J$ planet, all of the model core-accretion model planets are {\em substantially fainter} immediately after the end of accretion than the comparable hot start model at the same age.  A $10\rm\, M_J$ model is over two orders of magnitude fainter than if it experienced a hot start.  The difference in $L$ falls with mass,
 reaching a factor of two for a $2\,\rm M_J$ model.  The $1\,\rm M_J$ planet formed by
 core accretion is a factor of two brighter than produced by the equivalent hot start.

In Figure 4 we set time $t=0$ for the hot-start evolution to coincide with the first post-formation core accretion model. In this case, which maximizes the difference between the two approaches, the hot-start luminosity is larger for every planet mass, although the difference is again least for the lowest mass case.  As illustrated in this figure, the lower initial entropy of the core accreted planets manifests
as both a smaller initial radius and a much smaller effective temperature, both of which lead to a smaller luminosity. 
The hot start evolution predicts that the most massive models
at 1 Myr have a radius over twice that of Jupiter's and an effective temperature exceeding 2000 K.  By contrast,
the core accretion calculation predicts $R<1.5\,\rm R_J$ and  $T_{\rm eff} < 900\,\rm K$ for all cases.

Note that as the post-core accretion luminosity falls very slowly, the curves almost seem flat on the log-log plot.  This is because the small, cool, core accretion planets cool far more slowly than the large, bright, hot-start planets (see Eq. 2).  

A test of giant planet formation models is provided by the transiting hot Jupiters.  It is commonly postulated that the evolution of these planets is retarded when they arrive close to their parent star, since their thermal emission and atmospheric structure are dominated by the vast incident radiation (see the review by \citet{Charb06}). Given its  anomalously large radius of $1.320\pm0.025 \,\rm R_{Jup}$ \citep{Knutson06} for its mass of $0.66\pm 0.06\,\rm M_{Jup}$, some have suggested that an additional source of energy (also related to the proximity of the primary) may be helping to delay the contraction of the transiting
planet HD 209458 b.  Regardless of whether such a source 
exists--and assuming that the planet never grew in size--the large radius sets a lower limit to acceptable post-accretion radii for this planet.  As Figure 4 shows, the post-formation core accretion radius increases with falling mass and an isolated $1\,\rm M_J$ planet exceeds the observed radius of HD209458 b for over $10^7$ years, allowing plenty of time  for the planet to migrate to its current position \citep{Papal06}.  Thus the radius HD209458 b seems to be consistent with our core accretion model (although we did not compute a model for this precise mass).  Furthermore, our implementation of core accretion predicts that all non-accreting planets have radii less than $\sim1.4\,\rm R_J$, since the model radii of massive core-accreted planets are never larger than this value and their evolution cannot be halted at a larger size.  In contrast, a migrating, massive hot start planet could conceivably have a radius in excess of $1.6\,\rm R_J$ if its evolution were retarded early enough.

\subsection{Subsequent Evolution}

If infant core accreted Jupiters are fainter than their hot start cousins, for how long does the disparity persist?  
Figure 4 shows that by $10^7\,\rm yr$ the luminosity of a core accreted Jupiter is essentially identical to that of a hot start planet.  Since the initial luminosity disparity is greater with increasing mass, it is no surprise that more massive
core accreted planets take longer to match the hot start prediction.  However, the timescale
required is much larger than generally appreciated.  A $2\,\rm M_J$ planet takes almost $10^8$ year to match
the hot start track.  A  $10\,\rm M_J$ planet requires a full $10^9$ year to overcome its initial luminosity deficit.

\citet{Baraffe02} also evaluated the uncertainty in the early evolution of brown dwarfs and giant planets by comparing what might be termed  `hot start' and  `hotter start'
models.  Both are comparable to our `hot start' case, for example for their $5\,\rm M_J$ evolution the initial effective temperature was greater than 2000 K for both cases.  Not surprisingly (see Eq. 2), they found that their hotter models with even larger initial radii cooled very quickly and joined their `hot start' evolution tracks within a million years for all masses considered ($>5\,\rm M_J$).  This was the basis for their expressed confidence that the theoretical evolution tracks can be trusted for ages greater than a few million years.  Motivated by a preliminary report of our work \citep{YJAF05}, \citet{Chabrier06} also briefly considered the early evolution of cool 1 and $4\,\rm M_J$ planets with small initial radii.  They also found that smaller, cooler planets can take in excess of $10^7$ years to reach the standard hot-start luminosity tracks.

To illustrate the extreme sensitivity of the early cooling rate on initial entropy, we computed the Kelvin-Helmholtz cooling times for a $4\,\rm M_J$ planet with a variety of initial internal entropies.  For $S$ (expressed in units of $\rm k_B/baryon$) between 6 and 11, a range that more than spans the plausible initial entropies shown in Figure 2, we found that $\tau_{\rm HK}\propto e^{-2.8S}$.  Thus small changes in the initial estimate of $S$ produce disproportionately large changes the initial cooling rate.  Large values of $S$ yield fast cooling rates, and the rapid cooling rates led to the  conventional wisdom that planets rapidly forget their initial conditions.  But smaller values of $S$, which our implementation of the core accretion model predicts, yield much slower early cooling times and planets with longer memories.

For this reason, as we noted in Section 2.2, we do not place high confidence in the comparison between individual early evolution tracks for core accreted planets of different masses, because they differ from each other relatively little in initial $S$. {\em The 
details of the accretion shock and the mass accretion rate and timescale, which are essentially unknown, control $S$ and the relative initial post-accretion luminosity to a much greater degree than previously recognized}.  
In any case, for masses $\gtrsim 4\,\rm M_J
$, there is relatively little difference in the luminosity among the core accreted giant planets until about 30 to 50 million years after formation.  As a whole, our core accreted planets have substantially lower initial entropies and thus longer evolution times than
the hot start models. We plan to explore the early post-formation luminosity evolution of the core accreted planets, including the effects on satellites, in more detail in a future publication.

\subsection{Sensitivity to Assumptions}

To understand the sensitivity of the results to the limiting gas accretion rate we varied both the maximum gas accretion rate and the timescale for accretion cutoff during accretion phases 4 and 5, for a $2\,\rm M_J$ planet.  Results
are shown in Figure 5.  In the rapid gas accretion case, where the limiting accretion rate is set at $10^{-1}\,\rm {M_\earth}\,yr^{-1}$, the
final planet is formed very quickly, in less than $10^5$ yr after the start of runaway gas accretion.  The resulting planet is somewhat
larger and warmer than the baseline model.  Likewise a model with a very low accretion rate,   $10^{-3}\,\rm {M_\earth}\,yr^{-1}$,
and a long accretional tail off ends up cooler and smaller than the baseline case.  By 5 Myr, however, the differences between these cases are slight, less than 20\% in the total luminosity, much less than the factor of 2 difference
between the core accretion and hot start models for this mass.  We also considered a case for $10\,\rm M_J$ where we shortened, by a factor of 3, the time scale over which the accretion rate is linearly decreased from the limiting rate to zero.  Except for reaching the final mass more rapidly, this model behaved identically to the standard case and is not shown. 

Varying the temperature of the nebular gas, from $150\,\rm K$ to a very high $300\,\rm K$, delays the onset of runaway gas accretion and alters the final thermal state of the planet, but the luminosity difference is within the range found by only varying the mass accretion rate.  As shown in Figure 2, these relatively large changes in the nebular conditions for core accretion mechanism produce only slight differences in the initial entropy of the planet, which is the quantity that controls the subsequent evolution.

\subsection{Masses of Young Giant Planets}

Since young giant planets of known ages and dynamically-constrained masses have not yet been directly detected, the predictions of our model cannot yet be tested.  However young, planetary mass companions are now being discovered and characterized. We find it is illuminating to explore some of the properties of these objects--although they almost certainly formed by fragmentation and not core accretion--in light of our new understanding of the sensitivity of the evolution models to initial conditions.

The companion to 2MASS1207 \citep{Chauvin05, Song06} has an estimated luminosity of 
$\log (L\rm / L_\odot)\approx -4.30$ at an age of $8\pm3\,\rm Myr$.  Judging from the hot start evolution tracks shown in Figure 6, a mass of  about 3 to $7\,\rm M_J$ is reasonably inferred, consistent with the $5\pm3\,\rm M_J$ reported by \citet{Song06}. 
However, as we have seen, at such a young age, the model luminosities are highly dependent on the 
initial entropy of the evolution tracks.  Without a model connecting the formation process of these objects to their
initial entropy, the model-dependent uncertainty in their masses is unconstrained.

To illustrate this point, we computed (Fig. 6) two evolution tracks, for 4 and $10\,\rm M_J$ planets, with initial entropies
between the hot start and our core accretion cases.  The initial
conditions were chosen such that the entropy at 1 Myr would be equal to the mean the other two cases at an age of 1 Myr.  
Clearly if 2MASS1207b experienced such a `warm start', the derived mass would be closer to $8\rm \, M_J$ than to  $5\rm \, M_J$. While we make no claim to the applicability of such an arbitrary model to this particular object, we stress that until the model evolutionary curves can be calibrated at young ages, the derived masses are highly uncertain.  Even these intermediate `warm start' models take 
20 to 100 Myr to reach the standard model curves, a time span that still is  substantially longer than the age of many young clusters.

This point is further illustrated by a second putatitive planetary mass object, \object{GQ Lup b} \citep{Neuhauser05}.
Judging by our baseline hot-start luminosity tracks,  this object has a mass in excess of $10\,\rm M_J$.  
Lower initial model entropies would result in higher estimated masses.
Using a different set of models, with presumably higher initial entropies, the discoverers claimed a lower mass limit of $1\,\rm M_J$.  \citet{Neuhauser05} arrived at their low mass estimate by relying on evolution models
of \citet{Wuchterl03} that attempt to connect the initial conditions to the formation process, which is clearly a topic that
requires more attention.  

Indeed, the question of what is the proper initial condition to use in evolution models of giant planets and stars is an old one. \citet{Bodenheimer74}
recognized that the choice of a giant planet's final state after accretion would affect subsequent evolution.  Observations of the thermal emission of young planets with dynamically constrained masses and known ages will shed light on the nature of the giant planet formation process, particularly the role of the accretion shock.

\section{Conclusions}

We have computed the first giant planet evolution models that couple planetary thermal evolution to the predicted core mass and thermal structure of a core accretion planet formation model.   \citet{Baraffe06} investigated the evolution of planets with core sizes and heavy element abundances derived from the core accretion models of \citet{Alibert05}.  However, \citet{Baraffe06} did not attempt to match the thermal structure (and hence, temperature, entropy, and density) at the interface between planetary formation and subsequent evolution.

Our implementation of the core accretion model processes most of the planetary mass through an accretion shock in which
the accreting gas loses most of its internal entropy.  As a result, our young giant planets are cooler, smaller, fainter, and take longer to evolve than the standard hot-start model giant planets.  We note, however, that our accretion model  does not resolve the radiative transfer within the shock, but rather uses the  shock boundary conditions of \citet{stahler80}.  A more complete or detailed treatment of accretion at the surface of the planet could very well result in different initial conditions, including possibly a warmer, larger, brighter and more conventional young planets. Specifically, a three-dimensional hydrodynamic simulation of gas accretion by giant planets, allowing for material accreted through a circumplanetary disk and shock radiation, would provide more rigorous post-formation
models for subsequent evolution calculations.  Until such models are available, our approach--which likely provides a lower limit to the post accretion luminosity--demonstrates that plausible initial conditions can lead to very different early evolution tracks for giant planets than the `hot start' models that are commonly relied upon.

For example, at $10^7$ years--a time greater than the age of the TW Hydrae association--our $10\,\rm M_J$ core accreted planet is more than a factor of 100, or 5 magnitudes, fainter than the equivalent hot start planet.   The luminosity difference falls with decreasing mass, so that our model luminosity for a $1\,\rm M_J$ planet is comparable to the standard case.  Thus the thermal luminosity of young, massive giant planets, which have been assumed to be easy targets for coronagraphy, may be substantially less than previously assumed. 
If this result is correct, then searches for the thermal emission from young, several Jupiter mass planets must be
far more sensitive than previously anticipated in order to detect these relatively faint, young planets.  Thermal infrared planet searches by the Large Binocular Telescope, the James Webb Space Telescope, and other planned telescopes would all be impacted, although efforts to detect planets in reflected light would not.  This conclusion holds true even to ages as great as that of the Pleiades for the most massive planets considered here. 
Ironically the least massive, most intrinsically faint planets (1 to $2\,\rm M_J$) match their hot-start luminosity tracks by just a few tens of millions of years, or less and are much less underluminous before that time.   Direct detections of young giant planets with dynamically measured masses will test this conclusion.

Since the numerical values for luminosity that we derive depend primarily upon our treatment of energy radiated from the accretion shock,  these results should not  be viewed as specific predictions of the core accretion model.  Rather our point is that core accretion naturally leads to gas
accretion through a shock, which may result in low entropy planets.  The viability of giant planet formation via core accretion depends on physical processes happening earlier in the accretionary process (at smaller masses) than those processes that we have shown
to be crucial for the luminosity of young planets of Jupiter's mass and larger.

We note in passing that the faintness predicted for young Jupiter-mass planets compounds with non-equilibrium chemistry to make detection
of young giant planets at M band particularly challenging.  \citet{Marley96}, after the discovery of \object{Gl229B}, suggested that a substantial M-band 4 - 5 $\mu$m flux peak should be a universal feature of giant planets and brown dwarfs.  In addition to the intrinsic emergent flux, this spectral range has looked promising for planet detection due to the favorable planet/star flux ratio \citep[e.g.][]{Burrows05}.  
However, it has been known since the 1970s \citep[see][]{Prinn77} that Jupiter's 5 $\mu$m flux is suppressed by absorption by CO present in amounts exceeding that predicted by equilibrium chemistry.  This same effect has now been observed in brown dwarf M-band photometry \citep{Golim04, Leggett06}, as anticipated by \citet{Fegley96}.  Excess CO leads to strong absorption at 4.5 $\mu$m, leading to diminished flux in M-band \citep{Saumon03}.  This effect further suppresses the M band fluxes of young planets below the existing models.  Taken together, fainter young planets and reduced M-band flux may well reduce the catch from what had seemed a promising fishing hole for direct planet detection.

We also conclude that the predicted evolution of giant planets objects, regardless of formation mechanism,
is far more sensitive to the
precise  conditions at the termination of accretion than has been previously recognized.  Most workers have assumed that evolving model planets `forget' their initial
conditions within $10^6$ \citep{Baraffe03b} to $10^8$ years \citep{stevenson82} of the first time step.  While 1 to 2 Jupiter mass planets
do have a short ($\sim10^7\,\rm yr$) memories, we have shown that more massive planets remember their
initial thermal state far longer.  The evolution time scale for young, hot planets depends exponentially on their initial entropy.  Until the initial thermal
state of young, low mass objects--even isolated brown dwarfs--is known with more certainty, the early evolution tracks must be regarded with some skepticism. Any effort to assign a mass to a very young putative giant planet must consider the uncertainties in these early evolutionary tracks.

\acknowledgments

We acknowledge support from the NASA Origins and Planetary Atmospheres Programs, from the NSF Astronomy and Astrophysics Program, and 
from a NASA Postdoctoral Program (NPP) Fellowship (J. J. F.). We thank Richard Freedman and Katharina Lodders for assistance with molecular opacities
and chemistry, respectively.  We benefited from helpful exchanges with G$\rm \ddot{u}$nther Wuchterl and a helpful review by the anonymous referee.  


\clearpage

\begin{figure}
\includegraphics[scale=0.8]{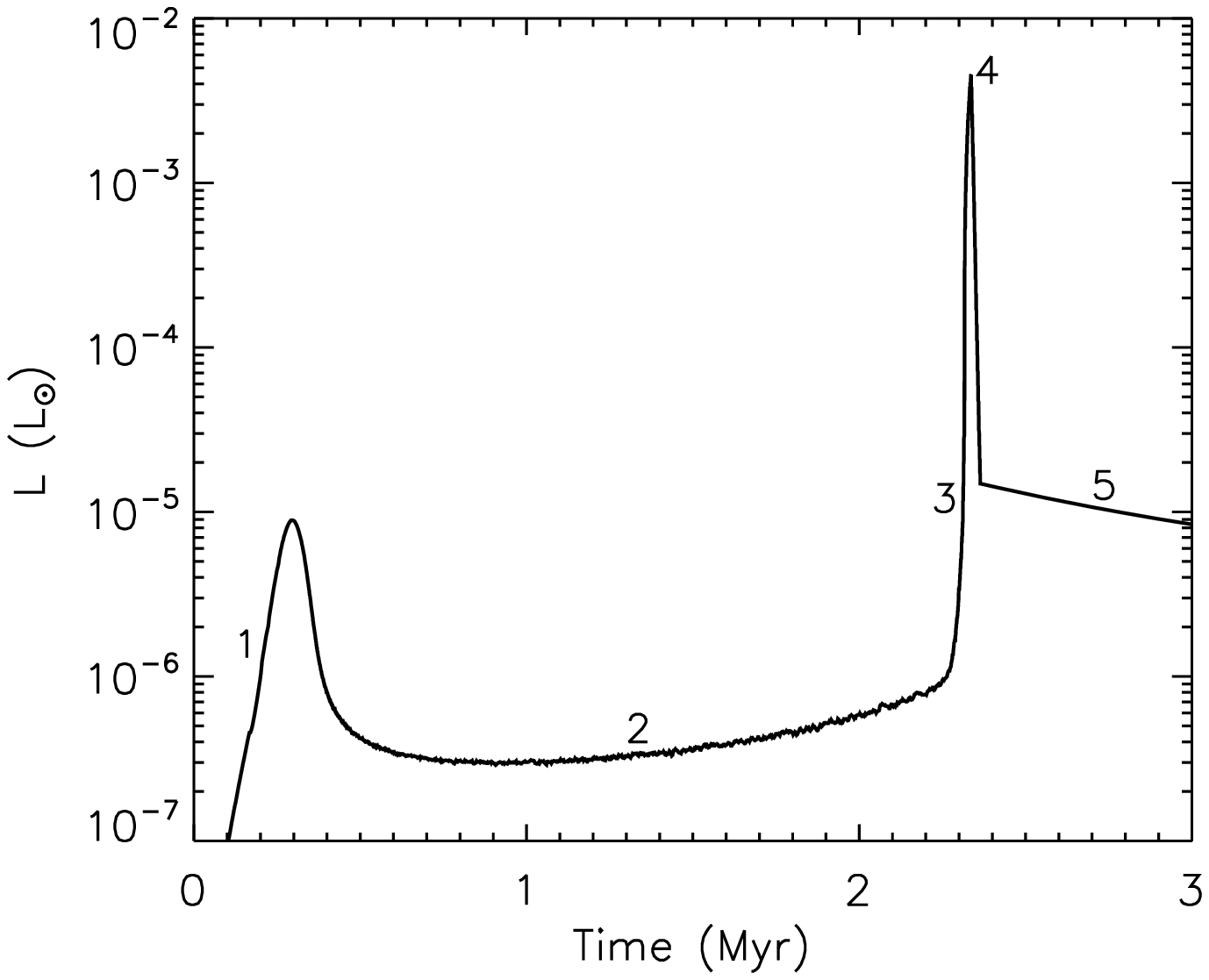}
\caption{Luminosity of a $1\,\rm M_J$ planet as a function of time. Numbers refer to various stages in the formation/contraction process as discussed in the text. In this figure, time $t=0$ is chosen to be the
start of the growth of the solid core. Model, through stage 4, is the  $10\rm L \infty$ case of Hubickyj et al. (2005). Subsequent evolution is calculated as described in Section 3. }
\end{figure}

\clearpage

\begin{figure}
\includegraphics{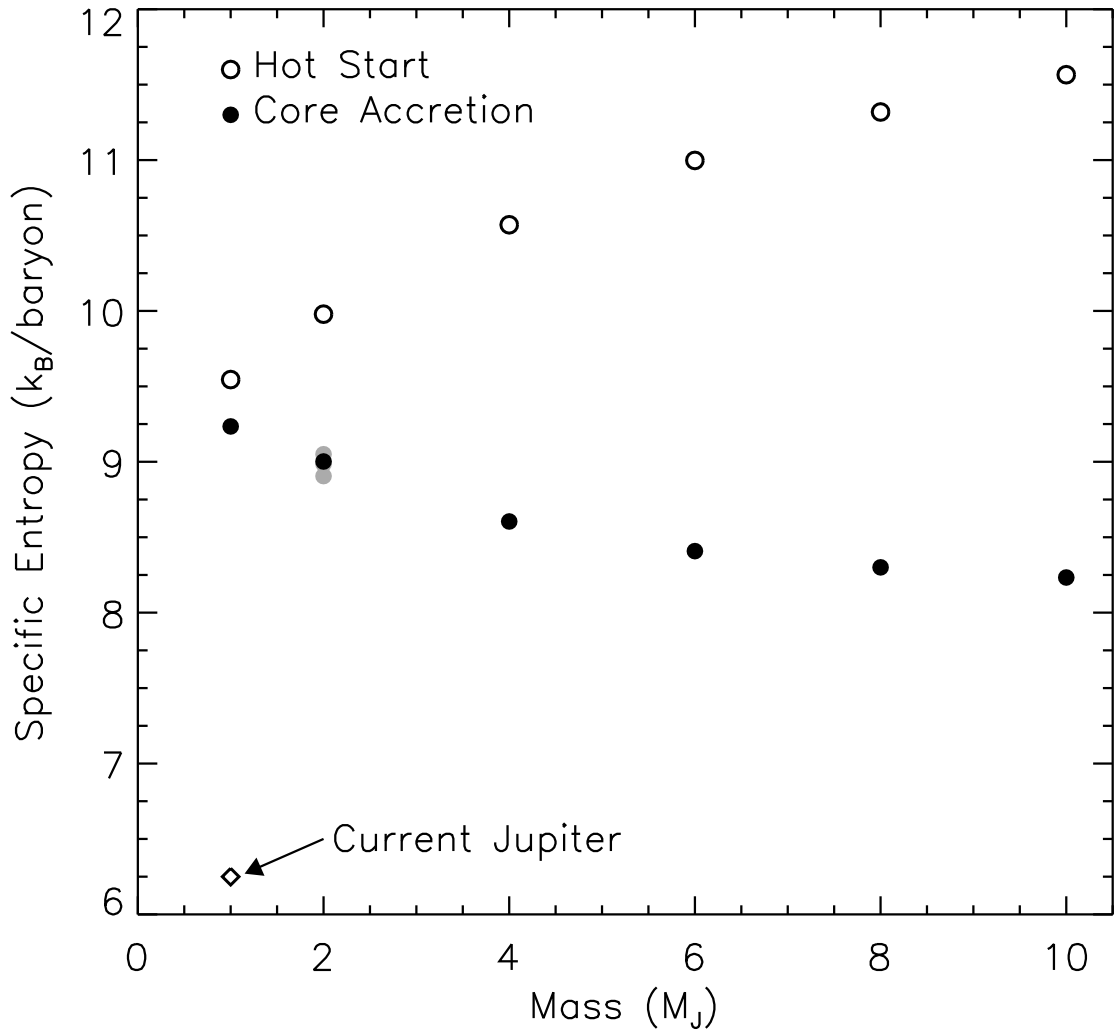}
\caption{Specific entropy of young giant planets formed by core accretion and the hot-start assumptions. Since
almost all of the mass of the planet sits on a single adiabat, the interior temperature-pressure conditions
can be characterized by the entropy of that adiabat.  For both cases the entropy plotted is at 1 million
years after the first time step in the evolution model.  Shaded circles at $2\,\rm M_J$ denote entropies of various alternate
cases for the core accretion model, as shown in Figure 5 and discussed in Section 4.3.  In the core accretion case this is 1 million years after the end of accretion. The entropy of the current Jupiter is also shown for comparison.}
\end{figure}

\clearpage

\begin{figure}
\includegraphics[scale=0.7,angle=90]{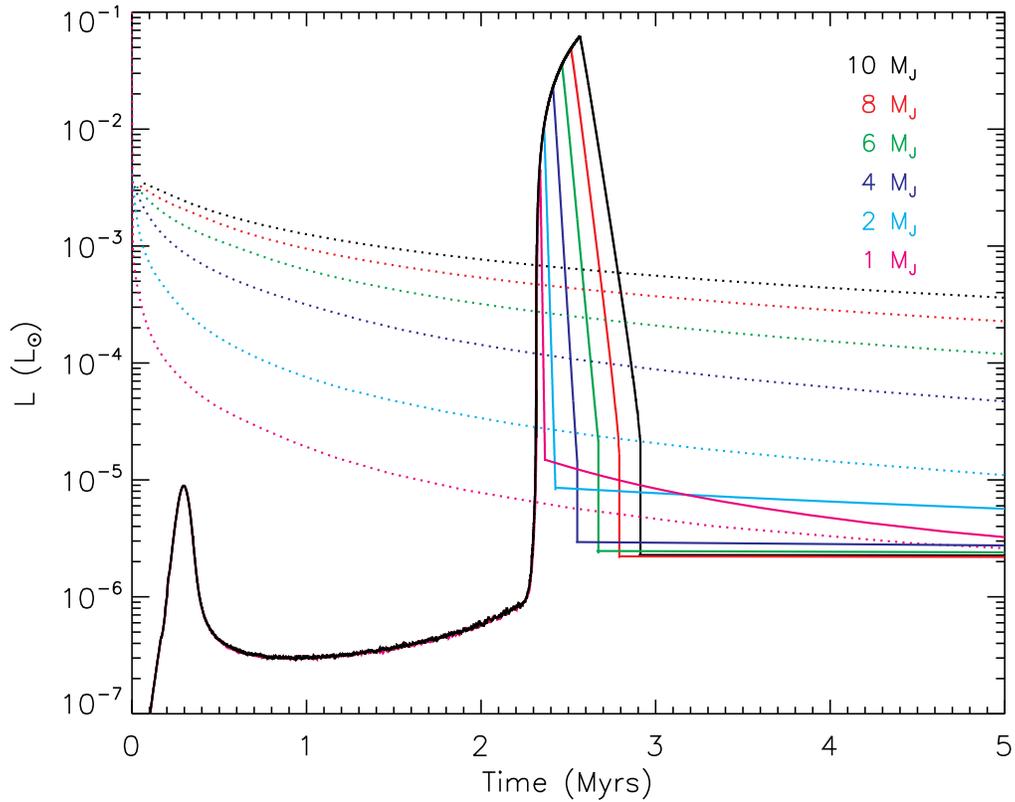}
\caption{Luminosity of young Jupiters of various masses as a function of time.  Dotted
lines are for a `hot start' evolution calculation as described in the text.  Solid 
lines denote the core accretion case. In this figure, time $t=0$ is chosen to be the
start of the growth of the solid core for the nucleated collapse scenario and the
first model of the `hot start' evolution. }
\end{figure}

\clearpage

\begin{figure}
\vspace*{-30mm}
\includegraphics[angle=90]{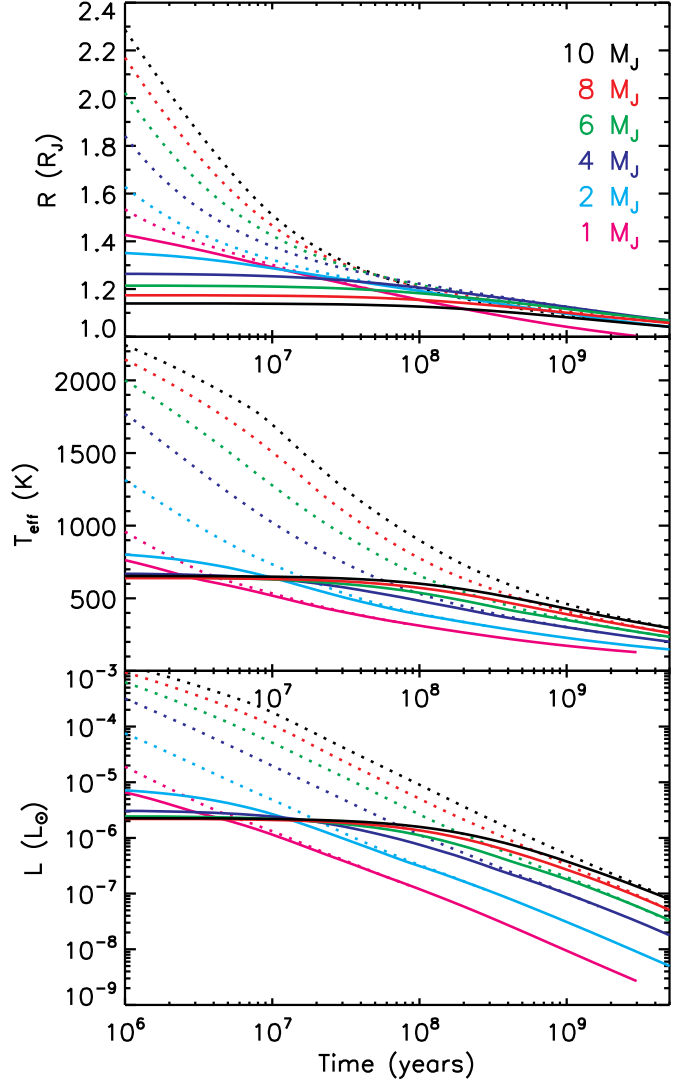}
\caption{Model radius, $R$, effective temperature, $T_{\rm eff}$, and
luminosity, $L$, of young Jupiters of various masses.  Line types as
in Figure 2.  Unlike Figure 2, in this figure time $t=0$ for the core
accretion evolution is chosen to be the last model of the core
accretion calculation.  There is thus an offset of 2.5 to 3 million
years, depending on mass, from Figure 2.  Both the radii and effective
temperature of the young planets are lower in the core accretion case,
leading to substantially lower luminosities.  Differences from the
`hot start' persist for as little as $10^7$ years for a $1\,\rm M_J$
planet to as much as $10^9$ years for a $10\,\rm M_J$ planet. }
\end{figure}

\clearpage

\begin{figure}
\includegraphics[scale=0.9]{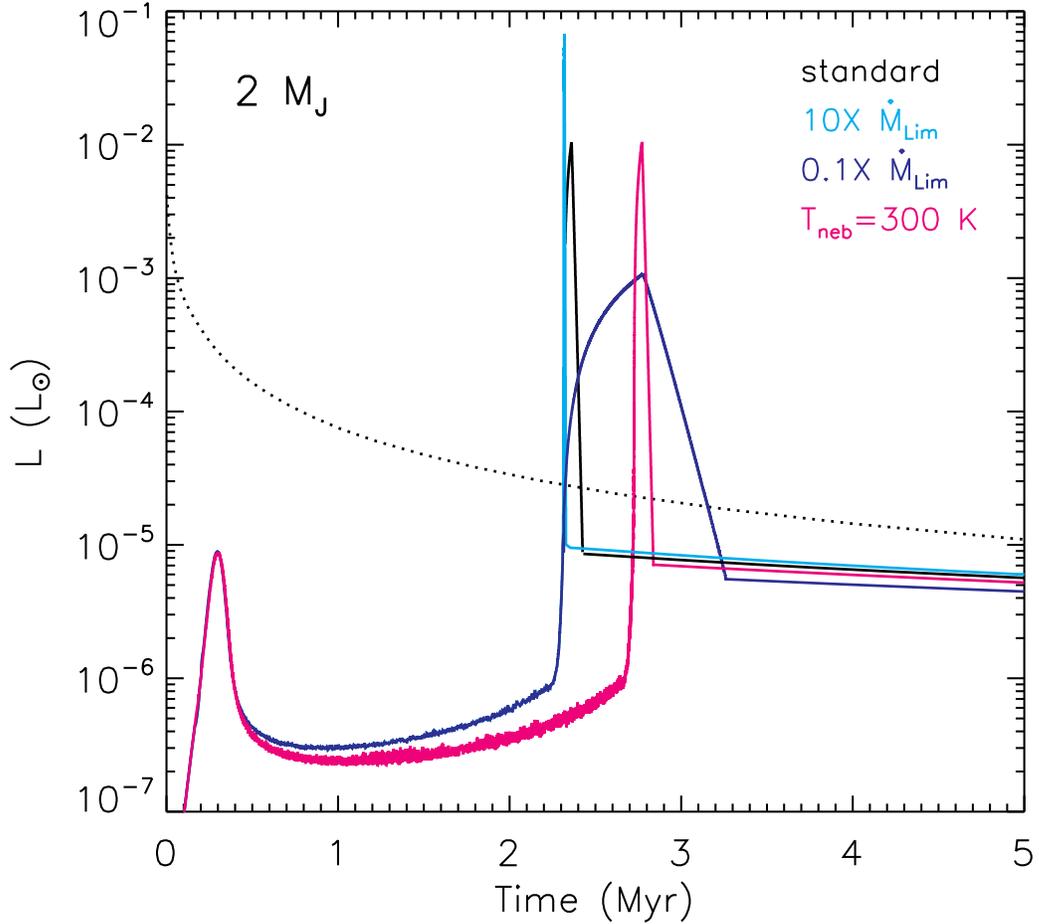}
\caption{Luminosity evolution of various $2\,\rm M_J$ cases discussed in the text. 
Black solid and dotted lines are the standard baseline core accretion and hot start models.  
Other line types are for 10 and 0.10 times the standard limiting mass accretion rate, 
${\dot M}_{\rm Lim}$, and a case with higher nebular gas temperature, $T_{\rm neb}$.}
\end{figure}

\clearpage

\begin{figure}
\includegraphics[scale=0.7]{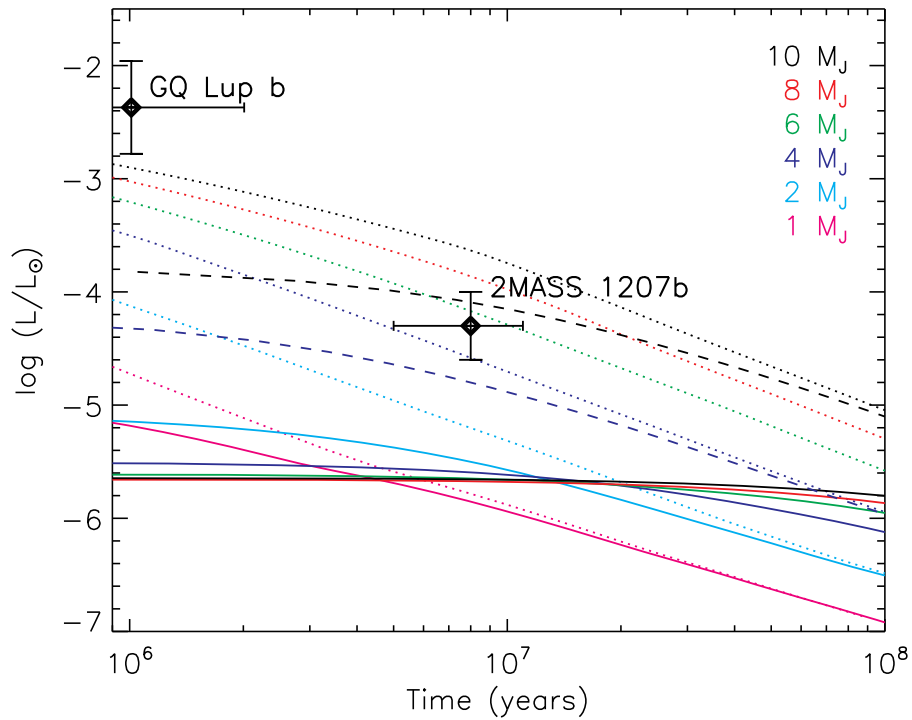}
\caption{ Luminosity evolution of various masses for the hot start (dotted), core accretion (solid), and intermediate
entropy (dashed) cases.   For the intermediate case, only tracks for the 4 and $10\,\rm M_J$ planets
are shown.   Also shown are the
estimated (model dependent) bolometric luminosities of two claimed \citep{Chauvin05, Neuhauser05} giant planet mass companions to more massive objects. }
\end{figure}

\end{document}